\documentclass[aps,prd,groupedaddress,amssymb,twocolumn,eqsecnum,showpacs,epsfig,nofootinbib]{revtex4}

\usepackage{graphicx}
\usepackage{bm}
\usepackage{dcolumn}
\usepackage{amsmath}
  
\usepackage{amsmath}
\usepackage{amssymb} 

\numberwithin{equation}{section}
\usepackage{epsfig}

\begin{document}
\newcommand{\newc}{\newcommand}

\newc{\be}{\begin{equation}}
\newc{\ee}{\end{equation}}
\newc{\bear}{\begin{eqnarray}}
\newc{\eear}{\end{eqnarray}}
\newc{\bea}{\begin{eqnarray*}}
\newc{\eea}{\end{eqnarray*}}
\newc{\D}{\partial}
\newc{\ie}{{\it i.e.} }
\newc{\eg}{{\it e.g.} }
\newc{\etc}{{\it etc.} }
{\newc{\etal}{{\it et al.}}
\newc{\lcdm}{$\Lambda$CDM}
\newcommand{\nn}{\nonumber}
\newc{\ra}{\rightarrow}
\newc{\lra}{\leftrightarrow}
\newc{\lsim}{\buildrel{<}\over{\sim}}
\newc{\gsim}{\buildrel{>}\over{\sim}}
\newcommand{\mincir}{\raise
-3.truept\hbox{\rlap{\hbox{$\sim$}}\raise4.truept\hbox{$<$}\ }}
\newcommand{\magcir}{\raise
-3.truept\hbox{\rlap{\hbox{$\sim$}}\raise4.truept\hbox{$>$}\ }}

%\title{The $\Lambda$CDM growth index revised}

\title{The $\Lambda$CDM growth rate of structure revisited}

%\title{A generalized growth index parametrization: Applications to the 
%$\Lambda$CDM using the {\em WiggleZ} growth data}

\author{Spyros Basilakos}\email{svasil@academyofathens.gr}
\affiliation{Academy of Athens, Research Center for Astronomy and
Applied Mathematics,
 Soranou Efesiou 4, 11527, Athens, Greece}

\begin{abstract}
We re-examine the growth index of the concordance 
$\Lambda$ cosmology 
in the light of the latest 6dF and {\em WiggleZ} data.
In particular, we investigate five different models for the growth 
index $\gamma$, by comparing their cosmological
evolution using observational data of the growth rate of structure formation 
at different redshifts.
Performing a joint likelihood analysis of 
the recent supernovae type Ia data, the Cosmic Microwave Background
shift parameter, Baryonic Acoustic Oscillations and 
the growth rate data, we 
determine the free parameters of the $\gamma(z)$ parametrizations 
and we statistically quantify their ability to represent the observations. 
We find that the addition of the 6dF and {\em WiggleZ} growth data  
in the likelihood analysis improves significantly 
the statistical results. 
As an example, considering a constant growth index 
we find $\Omega_{m0}=0.273\pm 0.011$ and
$\gamma=0.586^{+0.079}_{-0.074}$.

\end{abstract}
\maketitle

\section{Introduction}
The high-quality cosmological observational data (e.g. supernovae type
Ia, CMB, galaxy clustering, etc), accumulated during the last two decades,
have enabled cosmologists to gain substantial confidence that modern
cosmology is capable of quantitatively reproducing the details of
many observed cosmic phenomena, including the late time
accelerating stage of the Universe. A variety of studies
have converged to a cosmic expansion history involving a spatially
flat geometry and a cosmic dark sector formed by cold dark matter and some
sort of dark energy, endowed with large negative pressure, in order to
explain the observed accelerating expansion of the Universe
\cite{Teg04,Spergel07,essence,Kowal08,Hic09,komatsu08,LJC09,BasPli10,komatsu11}
(and references therein).

In spite of that, the absence of a fundamental physical theory, regarding
the mechanism inducing the cosmic acceleration, has given rise to a
plethora of alternative cosmological scenarios. 
Most are based either on the existence of new fields in nature (dark
energy) or in some modification of Einstein's general relativity,
with the present accelerating stage appearing as a sort of geometric effect.
In order to test the latter possibilities,
it has been proposed that measuring
the so called growth index, $\gamma$, could provide an efficient
way to discriminate between modified gravity models and dark energy 
(hereafter DE) models which adhere to general relativity.
The accurate determination of the growth index is considered one of
the most fundamental tasks on the interface between Astronomy and
Cosmology. Its importance steams from the fact that 
there is only a weak dependence of $\gamma$ on
the equation of state parameter $w(z)$, as has been found in 
Linder \& Cahn \cite{Linder2007}, which 
implies that one can separate the background expansion history, $H(z)$,
constrained by a large body of cosmological data (SNIa, BAO, CMB), 
from the fluctuation growth history, given by $\gamma$.
Assuming a homogeneous dark energy,
it was theoretically shown that for DE models
within general relativity the growth index $\gamma$ is 
well approximated by
$\gamma \simeq \frac{3(w-1)}{6w-5}$ 
(see \cite{Silv94},\cite{Wang98},\cite{Linder2007},\cite{Nes08}), which
boils down to $\approx 6/11$ for the $\Lambda$CDM cosmology $w(z)=-1$.
Notice, that in the case of the
braneworld model of Dvali, Gabadadze \& Porrati \cite{DGP}
we have $\gamma \approx 11/16$ 
(see also \cite{Linder2007,Gong10,Wei08,Fu09}), while for
the $f(R)$ gravity models we have $\gamma \simeq 0.41-0.21z$ 
for $\Omega_{m0}=0.27$ \cite{Gann09,Tsu09,Moto10}.

From the observational viewpoint, indirect methods 
to measure $\gamma$ have also been developed (mostly using a 
constant $\gamma$), 
based either on the observed growth rate of clustering  
\cite{Nes08,Guzzo08,Port08,Gong10,Dos10}
providing a wide range of $\gamma$ values
$\gamma=(0.60-0.67)^{+0.40\; +0.20}_{-0.30 \; -0.17}$,
or on massive galaxy clusters Vikhlinin et al. \cite{Vik09} and 
Rapetti et al. \cite{Rap10} with the
latter study providing $\gamma=0.42^{+0.20}_{-0.16}$,
or even on the weak gravitational lensing \cite{Daniel10}.
Gaztanaga et al.,\cite{Gazt12} performed 
a cross-correlation analysis between                                            probes of weak gravitational lensing and 
redshift space distortions and found no evidence for deviations 
from general relativity. 
Also, Basilakos \& Pouri \cite{Por} and 
Hudson \& Turnbull \cite{Hud12} used 
the combination parameter, namely 
$f(z)\sigma_{8}(z)$ (recently appeared in the literature), of the growth 
rate of structure,
$f(z)$, and the redshift-dependent rms fluctuations of the linear
density field, $\sigma_8(z)$ to constrain the growth index.
The above authors found $\gamma=(0.602-0.619)\pm 0.05$.
With the next generation of surveys, based on {\em Euclid} and
  {\em BigBOSS},
we will be able to put strong constraints on $\gamma$ 
(see for example \cite{LLin,Bel12,DP11} and references therein) and thus
to test the  validity of general relativity on cosmological scales.

%However, a weak point here is the fact that 
%most of these methods use as priors other
%cosmological parameters, and thus the resulting $\gamma$ measurements are 
%somehow model dependent. 

In this article, we wish to test some basic functional forms of 
$\gamma(z)$ in the 
light of the 6dF and {\em WiggleZ} growth rate data. 
The structure of the paper is as follows. 
Initially in section 2, we briefly discuss the background 
cosmological equations.
The basic theoretical elements of
the growth index are presented in section 3, where 
we extend the original Polarski \& Gannouji 
method \cite{Pol}
for a general family of $\gamma(z)$ parametrizations.
Notice that the current theoretical approach 
does not treat the possibility of having inhomogeneous DE.
In section~4, a joint statistical analysis
based on the {\em Union 2} set of type Ia supernovae 
(SNIa; \cite{Ama11}), the shift 
parameter of the Cosmic
Microwave Background (CMB; \cite{komatsu11}),  
the observed
Baryonic Acoustic Oscillations (BAOs; \cite{Perc10}) and 
the observed linear growth rate of 
clustering, measured mainly from the 2dF, VVDS, 
SDSS, 6dF and {\em WiggleZ} redshifts catalogs, 
is used to constraint the growth index model
free parameters.
Finally, we draw our main conclusions in section 5.

%This implies that the current bias evolution model 
%can be used to put constraints on dark energy 
%models as well as to investigate possible departures from 
%general relativity. 

\section{The background evolution}
In this section, it will be assumed that the universe is a
self-gravitating fluid described by general relativity, and endowed
with a spatially flat homogeneous and isotropic geometry.  In
addition, we also consider that it is filled by non-relativistic
matter plus a DE component (or some effective mechanism
that simulates it), and whose equation of 
state (hereafter EoS), $p_{DE}=w(a)\rho_{DE}$,
is driving the present accelerating stage. Following standard lines,
the Hubble flow reads:
\begin{equation}
\frac{H^{2}(a)}{H_{0}^{2}}\equiv 
E^{2}(a)=
\Omega_{m0}a^{-3}+\Omega_{DE0}{\rm
e}^{3\int^{1}_{a} d{\rm lny}[1+w(y)]},  \label{nfe1}
\end{equation}
where $a(z)=1/(1+z)$ is the scale factor of the universe, 
$E(a)$ is the normalized Hubble flow, $\Omega_{m0}$ is the
dimensionless matter density at the present epoch,
$\Omega_{DE0}=1-\Omega_{m0}$ denotes the DE density parameter
and $w(a)$ its EoS parameter. On the other hand, we can express 
the EoS parameter in terms of $E(a)=H(a)/H_{0}$ 
\cite{Saini00} using the Friedmann equations as
\begin{equation}
\label{eos22} 
w(a)=\frac{-1-\frac{2}{3}a\frac{{d\rm lnE}}{da}}
{1-\Omega_{m}(a)}
\end{equation}
where
\be 
\label{ddomm}
\Omega_{m}(a)=\frac{\Omega_{m0}a^{-3}}{E^{2}(a)} \;.
\ee
Differentiating the latter and utilizing
Eq.~(\ref{eos22}) we find that
\be
\label{domm}
\frac{d\Omega_{m}}{da}=
\frac{3}{a}w(a)\Omega_{m}(a)\left[1-\Omega_{m}(a)\right]\;.
\ee
Since the exact nature of the DE has yet to be found, the
above DE EoS parameter encodes our ignorance regarding the
physical mechanism powering the late time cosmic acceleration.

The methodology described above can also be applied to the framework
of modified gravity (see \cite{Linjen03, Linder2004}). In this case,
instead of using the exact Hubble flow through a modification of the
Friedmann equation one may consider an equivalent Hubble flow
somewhat mimicking  Eq. (\ref{nfe1}). The key point here is that
the accelerating expansion can be attributed to a kind of
``geometrical'' DE contribution. Now, since the matter
density (baryonic+dark) cannot accelerate the cosmic expansion, we
perform the following parametrization \cite{Linjen03, Linder2004}:
\begin{equation}
E^{2}(a)=\frac{H^{2}(a)}{H_{0}^{2}}= \Omega_{m0}a^{-3}+\Delta H^{2}.
\label{nfe2}
\end{equation}
Naturally, any modification to the Friedmann equation of general
relativity may be included in the last term of the above expression.
After some algebra one may also derive, using Eqs. (\ref{eos22}) and
(\ref{nfe2}), an effective (``geometrical'') dark energy EoS
parameter, given by:
\begin{equation}
\label{eos222} 
w(a)=-1-\frac{1}{3}\;\frac{d{\rm ln}\Delta
H^{2}}{d{\rm ln}a}.
\end{equation}
Notice that we will use the above quantities in the next section. 
%Section~\ref{sec:growth}.

%The above formulation will be adopted in our statistical analysis
%of all DE models discussed in section 4.

\section{The Evolution of the linear growth factor}\label{sec:growth}
Here, we briefly discuss
the basic equation which governs the behavior of the matter
perturbations on sub-horizon scales and within the framework of any 
DE model, 
including those of modified gravity (``geometrical dark energy''). 
At the su-horizon scales the DE component 
is expected to be smooth and thus it is
fair to consider perturbations only on the matter component of the
cosmic fluid \cite{Dave02}.
In the framework of the homogeneous DE, the evolution equation
of the matter fluctuations, for cosmological models where the DE
fluid has a vanishing anisotropic stress and the matter fluid is not
coupled to other matter species 
(see \cite{Lue04},\cite{Linder05},\cite{Stab06},\cite{Uzan07},\cite{Linder2007},\cite{Tsu08},\cite{Gann09},\cite{Dent}), is given by:
\be
\label{odedelta} 
\ddot{\delta}_{m}+ 2H\dot{\delta}_{m}=4 \pi G_{\rm eff} \rho_{m} \delta_{m} 
\ee
where $\rho_{m}$ is the matter density
and $G_{\rm eff}(t)=G_{N} Q(t)$, with $G_{N}$ denoting 
Newton's gravitational constant.

For those cosmological models which adhere to general relativity,
[$Q(t)=1$, $G_{\rm eff}=G_{N}$], the
above equation reduces to the usual time evolution
equation for the mass density contrast \cite{Peeb93}, while in the
case of modified gravity models (see \cite{Lue04},\cite{Linder2007},
\cite{Tsu08},\cite{Gann09}),
we have $G_{\rm eff}\ne G_{N}$ (or $Q(t) \ne 1$). 
In this context,
$\delta_{m}(t) \propto D(t)$, where $D(t)$ is the linear growing mode
(usually scaled to unity at the present time). 
%If we change the variables from $t$ to $a$
%($\frac{d}{dt}=H\frac{d}{d\ln a}$)
%then the time evolution
%of the mass density contrast (see Eq.~(\ref{odedelta})) takes the
%following form
%\be
%\label{dela}
%\frac{a^{2}}{\delta_{m}}\frac{d^{2}\delta_{m}}{da^{2}}+
%\left(3+a\frac{d{\rm ln}E}{da}\right)\frac{a}{\delta_{m}}
%\frac{d\delta_{m}}{da}=
%\frac{3}{2}\Omega_{m}(a)Q(a) \;.
%\ee

Solving Eq.(\ref{odedelta}) for
%Eq.(\ref{odedelta} or \ref{dela}) for 
the concordance $\Lambda$ cosmology\footnote{For the usual $\Lambda$CDM 
cosmological model we have 
$w(a)=-1$, $\Omega_{\Lambda}(a)=1-\Omega_{m}(a)$ and $Q(a)=1$.}, we
derive the well known
perturbation growth factor (see \cite{Peeb93}):
\be\label{eq24}
D(z)=\frac{5\Omega_{m0}
  E(z)}{2}\int^{+\infty}_{z}
\frac{(1+u)du}{E^{3}(u)} \;\;. 
\ee
In this work we use the above equation normalized to unity at the
present time.
Obviously, for $E(z) \simeq \Omega_{m0}^{1/2}\,(1+z)^{3/2}$ it
gives the standard result $D(z)\simeq a=(1+z)^{-1}$, which corresponds to the
matter dominated epoch, as expected.  

Now, for any type of DE,
an efficient parametrization
of the matter perturbations
is based on the growth rate of clustering
\cite{Peeb93}
\be
\label{fzz221}
f(a)=\frac{d\ln \delta_{m}}{d\ln a}\simeq \Omega^{\gamma}_{m}(a)
\ee
where $\gamma$ is the so called growth index
(see Refs.~\cite{Silv94,Wang98,Linjen03,Lue04,Linder2007,Nes08})
which plays a key role in cosmological studies as we described in the 
introduction, especially in the light of 
recent large redshift surveys 
(like the 6dF \cite{Beutler} and the 
{\em WiggleZ} \cite{Blake,Sam11}; and references therein).

\subsection{The generalized growth index parametrization}
Inserting the first equality of 
Eq.(\ref{fzz221}) into Eq.~(\ref{odedelta}) 
and using simultaneously
Eq.~(\ref{eos22}) and $\frac{d}{dt}=H\frac{d}{d\ln a}$, we derive after 
some algebra, that  
\be
\label{fzz222}
%3w(a)\Omega_{m}(a)[1-\Omega_{m}(a)]\frac{df}{d\Omega_{m}}+f^{2}+fX(a)  
%= \frac{3}{2}\Omega_{m}(a)Q(a) \;.
a\frac{df}{da}+f^{2}+X(a)f
= \frac{3}{2}\Omega_{m}(a)Q(a) \;,
\ee
where 
\be 
X(a)=\frac{1}{2}-\frac{3}{2}w(a)
\left[ 1-\Omega_{m}(a)\right] \;.
\ee
Now, we consider that the growth index varies with cosmic time.
Transforming equation (\ref{fzz222}) 
from $a$ to redshift [$\frac{d}{da}=-(1+z)^{-2}\frac{d}{dz}$]
and utilizing Eqs.(\ref{fzz221}) (\ref{domm}),
we simply derive the evolution equation 
of the growth index $\gamma=\gamma(z)$ (see also~\cite{Pol}).
Indeed this is given by:
\bear
\label{Poll}
&& -(1+z)\gamma^{\prime}{\rm ln}(\Omega_{m})+\Omega_{m}^{\gamma}+
3w(1-\Omega_{m})(\gamma-\frac{1}{2})+\frac{1}{2}\; \nonumber \\
&& =\frac{3}{2}Q\Omega_{m}^{1-\gamma} \;,
\eear
where prime denotes derivative with respect to redshift.
At the present epoch the above equation takes the form:
\bear
\label{Poll1}
&& -\gamma^{\prime}(0){\rm ln}(\Omega_{m0})+\Omega_{m0}^{\gamma(0)}+
3w_{0}(1-\Omega_{m0})[\gamma(0)-\frac{1}{2}]+\frac{1}{2}\; \nonumber \\
&&=\frac{3}{2}Q_{0}\Omega_{m0}^{1-\gamma(0)}\;,
\eear
where $Q_{0}=Q(z=0)$ and $w_{0}=w(z=0)$.

Over, the last few years there have been many theoretical 
speculations regarding the functional form of the growth index
and indeed various candidates have been proposed in the literature.
Here we phenomenologically parametrize $\gamma(z)$ 
by the following general relation
\be
\gamma(z)=\gamma_{0}+\gamma_{1}y(z)\;.
\ee
The latter equation can be seen as a first order Taylor expansion 
around some cosmological quantity such as $a(z)$, $z$ and $\Omega_{m}(z)$.
Interestingly, for those $y(z)$ functions which satisfy $y(0)=0$ 
[or $\gamma(0)=\gamma_{0}$] one can write the 
parameter $\gamma_{1}$ in terms of $\gamma_{0}$. 
In this case [$\gamma^{\prime}(0)=\gamma_{1}y^{\prime}(0)$], using 
Eq.(\ref{Poll1}) we obtain
\be
\label{Poll2}
\gamma_{1}=\frac{\Omega_{m0}^{\gamma_{0}}+3w_{0}(\gamma_{0}-\frac{1}{2})
(1-\Omega_{m0})-\frac{3}{2}Q_{0}\Omega_{m0}^{1-\gamma_{0}}+\frac{1}{2}  }
{y^{\prime}(0)\ln  \Omega_{m0}}\;.
\ee
Note that for the rest of the paper 
we concentrate on the usual $\Lambda$CDM cosmology and thus we  
set $Q(z)=1$.

Let us now briefly present various forms of $\gamma(z)$, $\forall z$.
\begin{itemize}

\item Constant growth index (hereafter $\Gamma_{0}$ model): Here
we set $\gamma_{1}$ strictly equal to zero, thus $\gamma=\gamma_{0}$.

\item Expansion around $z=0$ (see \cite{Pol}; hereafter $\Gamma_{1}$ model): 
In this case we have $y(z)=z$. Note however, that this parametrization is valid 
at relatively low redshifts $0\le z \le 0.5$. 
In the statistical 
analysis presented below we utilize a constant 
growth index, namely $\gamma=\gamma_{0}+0.5\gamma_{1}$ for $z>0.5$.

\item Interpolated parametrization 
(hereafter $\Gamma_{2}$ model): Since $\Gamma_{1}$ model
is valid at low redshifts we propose to use a new formula 
$y(z)=z{\rm e}^{-z}$ 
that connects smoothly low and high-redshifts ranges. 
The above formula can be viewed as a combination of $\Gamma_{1}$ model
with that of Dossett et al.\cite{Dos10}. 
%Notice, that 
%we define $z_{t}$ to be that
%redshift (order of unity) at which the growth rate of structure begins to
%transition to its asymptotic value $f(z)\sim 1$. 
For $z\gg 1$ we have 
$\gamma_{\infty}\simeq \gamma_{0}$.

\item Expansion around $a=1$ (\cite{Bel12,DP11,Ishak09}; hereafter 
$\Gamma_{3}$ model): Here 
the function $y$ becomes $y(z)=1-a(z)=\frac{z}{1+z}$. Obviously, 
at large redshifts $z\gg 1$ we get 
$\gamma_{\infty}\simeq \gamma_{0}+\gamma_{1}$.    

\item Expansion around $\Omega_{m}=1$ (\cite{Wang98}; hereafter $\Gamma_{4}$ 
model\footnote{Concerning the $\gamma_{1}$ parameter, 
Gong et al.\cite{Gong09} found a rather different value 
$\gamma_{1}=\frac{3}{125}\frac{(1-w_{0})(1-3w_{0}/2)}{(1-6w_{0}/5)^{2}(1-12w_{0}/5)}$ }): 
In this parametrization we have $y(z)=1-\Omega_{m}(z)$
implying that 
$y(0)=1-\Omega_{m0} \ne 0$. As we have alerady mentioned above, the 
latter condition means that we can not write $\gamma_{1}$ in terms 
of $\gamma_{0}$. However, considering   
a constant equation of state parameter $w(z)=w_{0}=const.$ one 
can write $(\gamma_{0},\gamma_{1})$ only in terms of 
$w_{0}$ \cite{Wang98,Gong09} 
%For a constant equation of state parameter $w(z)=w_{0}$ one 
%can write $(\gamma_{0},\gamma_{1})$ in terms of $w_{0}$
\be
\label{WA98}
\gamma_{0}=\frac{3(1-w_{0})}{5-6w_{0}} \;\;\;
\gamma_{1}=\frac{3}{125}\frac{(1-w_{0})(1-3w_{0}/2)}{(1-6w_{0}/5)^{3}} \;.
\ee
Since at large redshifts $\Omega_{m}\simeq 1$ we can write 
$\gamma_{\infty}\simeq \gamma_{0}$.   
\end{itemize}
To conclude, for the $\Gamma_{1}$, $\Gamma_{2}$ and $\Gamma_{3}$ 
parametrizations one can show that $y(0)=0$ and 
$y^{\prime}(0)=1$, respectively. 
Evidently, based on the above 
discussion it becomes clear that Eq.(\ref{Poll2}) is satisfied 
for any type of DE model. 
Therefore, for the case of the $\Lambda$CDM
cosmology with $\gamma_{0}\simeq 6/11$ and 
$\Omega_{m0}=0.274$, Eq.(\ref{Poll2}) provides $\gamma_{1}\simeq -0.0477$, 
while for the case of the $\Gamma_{4}$ model we obtain 
$\gamma_{1}\simeq 0.01127$ (see Eq.\ref{WA98}). 

\begin{table}[ht]
\caption[]{The growth data.
The correspondence of the columns is as follows: index, redshift, observed
growth rate and references.
In the final column one can find various
symbols of the data appearing in Fig.5.} \tabcolsep 4.5pt
\vspace{1mm}
\begin{tabular}{ccccc} \hline \hline
Index&$z$& $f_{obs}$ & Refs.& Symbols \\ \hline
1&0.15 & $0.49\pm 0.14$&\cite{Guzzo08,Verde02,Hawk03}&solid circles\\
2&0.35 & $0.70\pm 0.18$&\cite{Teg06} &solid circles\\
3&0.55 & $0.75\pm 0.18$&\cite{Ross07}&solid circles\\
4&0.77 & $0.91\pm 0.36$&\cite{Guzzo08}&solid circles\\
5&1.40 & $0.90\pm 0.24$&\cite{daAng08}&solid circles\\
6&2.42 & $0.74\pm 0.24$&\cite{Viel04,Dos10}&solid circles\\
7&3.00 & $1.46\pm 0.29$&\cite{McDon05}&solid circles\\
8&0.067& $0.58\pm 0.11$&\cite{Beutler}&open triangles\\
9&0.22 & $0.60\pm 0.10$&\cite{Blake}&open circles\\
10&0.41 & $0.70\pm 0.07$&\cite{Blake}&open circles\\
11&0.60 & $0.73\pm 0.07$&\cite{Blake}&open circles\\
12&0.78 & $0.70\pm 0.08$&\cite{Blake}&open circles\\ \hline\hline \label{tab:growth}
\end{tabular}
\end{table}

\section{Observational constraints}
In the following we briefly present some details of the
statistical method and on the observational sample 
that we adopt in order to constrain the free parameters of the growth index,
presented in the previous section. 

\subsection{The Growth data}
The growth data that we will use in this work based on
the 2dF, VVDS, SDSS, 6dF and {\em WiggleZ} galaxy surveys,
for which the observed growth rate of structure,
$f_{obs}(z)$, is provided as a function of redshift.
In Table 1 we quote the precise numerical values of the data points 
with the corresponding errors.
This is an expanded version of the data-set 
used in \cite{Nes08,Gong10,Bas11} in which we have included data from the 
6dF \cite{Beutler} and 
{\em WiggleZ} \cite{Blake} galaxy survey (see entries $8-11$ in Table I). 
I particular the data used are based on

\begin{itemize}
\item The 2dF (Verde et al. \cite{Verde02}; Hawkins et al. 
\cite{Hawk03}), SDSS-LRG 
(Tegmark et al. \cite{Teg06}), the combined catalog of 2dF/SDSS-LRG/2SLAQ
(Ross et al. \cite{Ross07}; da Angela et al. \cite{daAng08}), 
VVDS (Guzzo et al. \cite{Guzzo08}), the LUQAS quasi-stellar object 
sample (Viel et al. \cite{Viel04}) and the 
SDSS Ly-$\alpha$ forest (McDonald et al.\cite{McDon05}) 
growth results as collected by
\cite{Nes08,Gong10,Bas11}. This data-set contains 7 entries.
%Note that different authors used different
%fiducial $\Lambda$CDM background cosmologies (see the 
%discussion below and Table I). 

%In Table I, one may see a more compact presentation
%of the current growth data including the
%fiducial $\Lambda$CDM models.

\item The 6dF growth rate data (1 entry) of Beutler et al. \cite{Beutler}
based on the (6dFGS) survey which contains $\sim 81971$ 
low redshit galaxies ($z\le 0.18$).
 
\item The {\it WiggleZ} growth measurements (4 entries)
of Blake et al. \cite{Blake}
based on spectroscopic data of $\sim$152000 galaxies
in the redshift bin $0.1<z<0.9$.
%and on a $\Lambda$CDM fiducial cosmology
%with $\Omega_{m0}=0.27$ and $\sigma_{8}=0.81$. 
%Notice that $\sigma_{8}$ is the rms mass 
%fluctuation on $8 h^{-1}$ Mpc scales at redshift $z=0$. 

\end{itemize}
Interestingly, the first (old) sample 
measures the growth rate to within an uncertainty of $20-40\%$ while the 
latter two $9-17\%$.
The observed growth rate of structure ($f_{obs}=\beta b$) is derived 
from the redshift space 
distortion parameter $\beta(z)$ and the linear bias $b(z)$. 
Observationally, the distortion parameter is
measured by the redshift-space two-point correlation
function. However, the main caveat here 
is related with the linear bias factor which can be defined 
as the ratio of the variances of the tracer (galaxies, QSOs etc) 
and underlying mass
density fields, smoothed at $8h^{-1}$ Mpc 
$b(z)=\sigma_{8,tr}(z)/\sigma_{8}(z)$, where
$\sigma_{8,tr}(z)$ is measured directly from the sample. 

Therefore, the weak point of using the $f_{obs}(z)$ data is  
the fact that $\sigma_{8}(z)[=\sigma_{8}D(z)$] is 
defined using a particular fiducial (reference) $\Lambda$CDM 
model (see also \cite{Nes08}) 
%Obviously, the fiducial $\sigma_8$ value only moves 
%the input power up or down (due to $f_{obs} \propto \sigma_{8}$) 
%which implies that it is not really important 
%for the measured $f_{obs}(z)$. 
%In contrast, the assumed (fiducial) 
%value of $\Omega_{m0}$ affects in a complicated way 
%the growth measurements [via the growth factor $D(z)$] 
which means that the current $f_{obs}(z)$ data can only be used
to test consistency of $Lambda$CDM cosmology \cite{Nes08}.  
In order to alleviate the above problems 
Song \& Percival \cite{Song09} proposed another estimator, namely 
$f(z)\sigma_{8}(z)$ which is almost a model-independent
way of expressing
the observed growth history of the universe.
%makes the growth rate sample rather heterogeneous.
However, despite the above caveats the $f_{obs}(z)$ growth 
rate data have been used extensively 
in the literature in order to put constraints on the growth 
index $\gamma$ (see for example \cite{Nes08,Gong10,Wei08,Dos10}).

\begin{figure}[ht]
\includegraphics[width=0.5\textwidth]{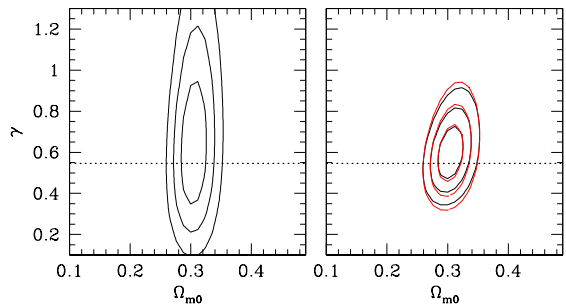}
\caption{{\em Left Panel:} Likelihood contours (for $-2{\rm ln}{\cal
L}_{tot}/{\cal L}_{max}$ equal to 2.30, 6.18 and 11.83, corresponding
to 1$\sigma$, 2$\sigma$ and $3\sigma$ confidence levels) in the
$(\Omega_{m0},\gamma)$ plane. Note that in 
the overall likelihood function we use the old growth data
(see entries 1-7 in Table I) and the expansion data (SNIa/CMB/BAO). 
{\em Right Panel:} Here we show the corresponding
contours based on the joint statistical analysis of the 
total growth+expansion (solid lines) and the 
6dF/{\em WiggleZ} growth+expansion (dashed lines) data respectively.
Note that the straight line corresponds 
to $\gamma=6/11$.}
\end{figure}

\subsection{The overall Likelihood analysis}
In order to constrain the cosmological parameters
of the concordance $\Lambda$CDM model 
one needs to perform a joint likelihood
analysis, involving the cosmic expansion data such as 
SNIa, BAO and CMB shift parameter together with the growth data.
Up to now, due to the large errors of the growth data with respect
to the cosmic expansion data, various authors preferred to 
constrain first $\Omega_{m0}$ using SNIa/BAO/CMB 
and then to fit $\Omega^{\gamma}_{m}(z)$ to the growth data $f_{obs}(z)$ alone. 
Of course, in the light of the 6dF and 
{\em WiggleZ} growth data it would be worthwhile
to simultaneously constrain the $(\Omega_{m0},\gamma)$ pair.
%In this section we attempt place constraints on
%the main cosmological parameters of the $\Lambda$CDM model by 
%performing a joint likelihood analysis utilizing 
%the growth data and the so called expansion rate data. 
In particular, we use the {\em Union 2} set of 557 SNIa of 
Amanullah et al.\,\cite{Ama11}\,\footnote{The expansion data and 
the corresponding covariances can be found in 
http://supernova.lbl.gov/Union/ and 
in the paper of Zhang et al. \cite{Zhang12}.}, the shift 
parameter of the CMB\,\cite{komatsu11} and the observed
BAOs (see \cite{Perc10}).
The overall likelihood function\footnote{Likelihoods are
normalized to their maximum values. In the present analysis we
always report $1\sigma$ uncertainties on the fitted parameters.
The total number of expansion data points used here is
$N_{E}=559$, while the associated degrees of freedom is: {\em
  dof}$= N_{E}+N_{f}-n_{\rm fit}-1$, where $N_{f}$ 
is the number of growth entries used in the 
statistical analysis and $n_{\rm fit}$ is
the model-dependent number of fitted
parameters. Note that the uncertainty of the fitted 
parameters will be estimated, in
the case of more than one such parameters, by marginalizing one with respect
to the others.} 
is given by the product of the individual
likelihoods according to:
\begin{equation}\label{eq:overalllikelihood} 
{\cal L}_{tot}({\bf p})={\cal L}_{E} (\Omega_{m0})
%{\cal L}_{SNIa}(\Omega_{m0}) \times {\cal L}_{CMB} (\Omega_{m0})
%\times {\cal L}_{BAO}(\Omega_{m0})
\times {\cal L}_{f}({\bf p})
\end{equation}
where
\begin{equation}\label{eq:overalllikelihood1} 
{\cal L}_{E} (\Omega_{m0})=
{\cal L}_{SNIa}\times {\cal L}_{CMB} 
\times {\cal L}_{BAO} \;. 
\end{equation}
Since likelihoods are defined as ${\cal L}_j\propto
\exp{\left(-\chi_j^2/2\right)}$, it translates into an addition for the
joint $\chi^2_{tot}$ function:
\begin{equation}\label{eq:overalllikelihoo}
\chi^{2}_{tot}({\bf p})=\chi^{2}_{E}(\Omega_{m0})+ 
\chi^{2}_{f}({\bf p})
\end{equation}
with 
\begin{equation}\label{eq:overalllikelihoo1}
\chi^{2}_{E}(\Omega_{m0})=
\chi^{2}_{SNIa}+ 
\chi^{2}_{CMB}+\chi^{2}_{BAO} \;.
\end{equation}
Note that the $\chi^{2}_{f}$ is given by
\be
\chi^{2}_{f}({\bf p})=\sum_{i=1}^{N_{f}} \left[ \frac{f_{obs}(z_{i})-
\Omega_{m}(z_{i})^{\gamma(z_{i},{\bf p})}}
{\sigma_{i}}\right]^{2} 
\ee
where $\sigma_{i}$ is the observed growth rate uncertainty.
Evidently, the essential free parameters that enter in 
Eq.(\ref{eq:overalllikelihoo}) are:
${\bf p}\equiv (\Omega_{m0}, \gamma_{0},\gamma_{1})$.

\begin{figure}[ht]
\includegraphics[width=0.5\textwidth]{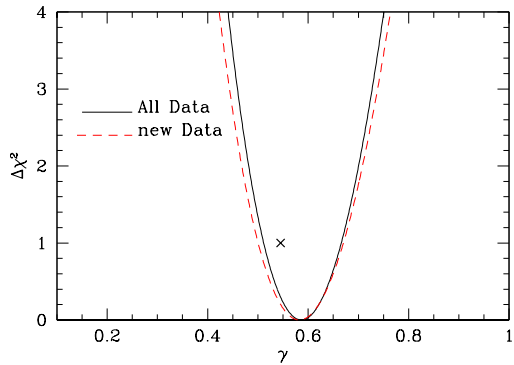}
\caption{The variance $\Delta \chi^{2}=\chi^{2}_{tot}-\chi^{2}_{min}$ 
around the best fit $\gamma$ value when we marginalize over
$\Omega_{m0}=0.273$. The solid and the 
dashed line correspond to the total sample (see Table I) and 
to the 6dF/{\em WiggleZ} data respectively. Note that the cross corresponds 
to $(\gamma,\Delta \chi_{1\sigma}^{2})=(6/11,1)$.}\label{fig:DEE}
\end{figure}

\subsubsection{Constant growth index}
%Here we set $\gamma_{1}$ strictly equal to zero which implies 
%$\gamma=\gamma_{0}$ (hereafter $\Gamma_){0}$ model). 
First of all we utilize the $\Gamma_{0}$ parametrization
($\gamma=\gamma_{0}$, $\gamma_{1}=0$: see section 3A).
Therefore, the corresponding statistical vector ${\bf p}$ contains
only two free parameters namely, ${\bf p}=(\Omega_{m0},\gamma,0)$. 
We sample $\Omega_{m0} \in [0.1,0.6]$ and
$\gamma\in [0.1,1.3]$ in steps of 0.001. 
In figure 1 we present the
1$\sigma$, 2$\sigma$ and $3\sigma$ confidence levels in the
$(\Omega_{m0},\gamma)$ plane for the old growth+SNIa/CMB/BAO 
(left panel) and the total growth+/SNIa/CMB/BAO data (right panel) respectively.
Using the old growth data (see entries 1-7 in Table I)
it is evident that although the $\Omega_{m0}$ parameter is tightly
constrained ($0.274\pm 0.012$), the $\gamma$ parameter remains 
weakly constrained, $\gamma=0.607^{+0.197}_{-0.174}$ with 
$\chi^{2}_{min}/dof\simeq 547.2/563$.
As can be seen in the right panel of figure 1, the strong 
$2\sigma$-degeneracy
is broken when we include the 6dF and the {\em WiggleZ} growth data in 
the joint likelihood analysis. Indeed 
using the total growth data-set (see solid line in the left panel of figure 1)
we find that the overall
likelihood function peaks at $\Omega_{m0}=0.273 \pm 0.011$ and
$\gamma=0.586^{+0.079}_{-0.074}$ ($\chi_{min}^{2}/dof
\simeq 549.3/568$), while using only the 6dF/{\em WiggleZ} data 
(dashed line; $N_{f}=5$, entries 8-12)
we obtain $\gamma=0.583^{+0.087}_{-0.082}$ ($\chi_{min}^{2}/dof
\simeq 545.3/567$). 
Furthermore, it becomes evident that 
using the overall growth data-set together with the expansion cosmological 
data we decrease the $2\sigma$ surface area (see the left panel of figure 1) 
by a factor of $\sim 2.5$. Hereafter we call this quantity ''reduction factor''
and is indicated by $S$ (closely related to 
the ''figure-of-merit'' definition), defined as the ratio of the surface area 
of the $2\sigma$ contour using the old growth data to that
of the total growth data-set.
It is also interesting to mention that the best fit value of 
$\Omega_{m0}$ is in excellent agreement with that provided 
by WMAP7 ($\Omega_{m0}=0.273$; Komatsu et al. \cite{komatsu11}).
In figure 2 we plot the variation of
$\Delta \chi^{2}=\chi^{2}_{tot}(\gamma)-\chi^{2}_{min}(\gamma)$
around the best $\gamma$ fit value 
when we marginalize over $\Omega_{m0}=0.273$.

\begin{figure}[ht]
\includegraphics[width=0.5\textwidth]{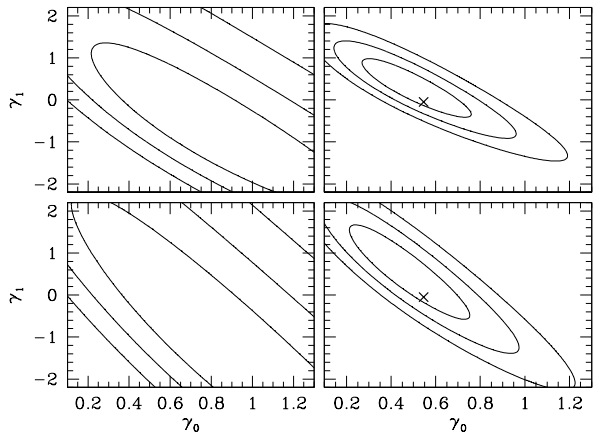}
\caption{Likelihood contours (for 
$\Delta \chi^{2}=-2{\rm ln}{\cal L}/{\cal L}_{\rm max}$ equal to 2.30, 
6.18 and 11.83, corresponding
to 1$\sigma$, 2$\sigma$ and $3\sigma$ confidence levels) in the
$(\gamma_{0},\gamma_{1})$ plane in the case of 
$\Gamma_{1}$ ({\em upper panel}) and $\Gamma_{2}$ ({\em bottom panel}) 
parametrizations (see section 3A).
In the left panels we 
present the contours that correspond to the old growth rate data 
(see Table I, entries 1-7) while the {\em right panels} show the likelihood
contours for the overall sample including the 6dF/{\em WiggleZ} data.
We also
include the theoretical $\Lambda$CDM ($\Omega_{m0}=0.273$; crosses) pair 
$(\gamma_{0},\gamma_{1})=(6/11,-0.0477)$. 
}\label{fig:contours}
\end{figure}

The above $\gamma$ best fit results are in agreement 
(within $1\sigma$) with
%and almost $1\sigma$ ($\Delta \chi_{1\sigma}^{2}=1$) away, 
the theoretically predicted value of $\gamma \simeq 6/11$ 
(see straight line in figure 1). Also our growth index results are in agreement
with previous studies.
%Such a discrepancy between the theoretical 
%$\Lambda$CDM and observationally fitted
%value of $\gamma$ has also been found
%by other authors. 
For example, Di Porto \& Amendola
\cite{Port08} obtained $\gamma=0.60^{+0.40}_{-0.30}$, 
Gong \cite{Gong10} measured $\gamma=0.64^{+0.17}_{-0.15}$ 
while Nesseris \& Perivolaropoulos \cite{Nes08},
found $\gamma=0.67^{+0.20}_{-0.17}$.
Comparing the error bars among the various best fit values 
it is interesting to mention that including 
in the likelihood analysis the 6dF and the {\em WiggleZ} data we manage 
to reduce the error budget by $\sim 50\%$.
Finally it is interesting to mention that our $\gamma$ parameter is 
in excellent agreement with those found based on the $f\sigma_{8}$ estimator.
Indeed Samushia et al. \cite{Sam11} found $\gamma=0.584\pm 0.112$, 
Rapetti et al. \cite{Rap12} obtained $\gamma=0.576^{+0.058}_{-0.059}$, 
Basilakos \& Pouri \cite{Por} and Hudson \& Turnbull \cite{Hud12}
found $\gamma=0.602\pm 0.055$ and $\gamma=0.619\pm 0.054$ 
respectively. 

\begin{figure}[ht]
\includegraphics[width=0.5\textwidth]{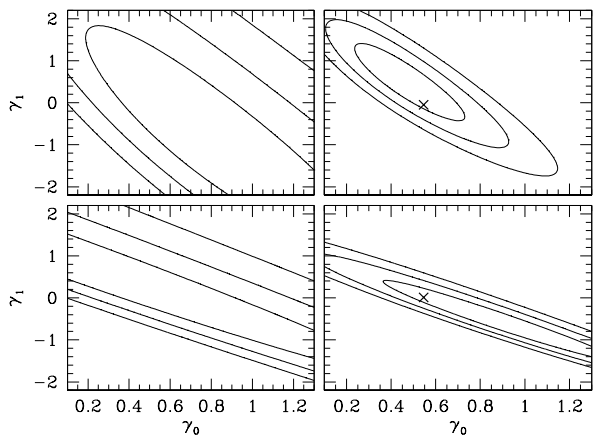}
\caption{The Likelihood contours for
$\Gamma_{3}$ ({\em upper panel}) and $\Gamma_{4}$ ({\em bottom panel}).
For more definitions see caption of figure 3.
Here the crosses correspond to the theoretical 
$(\gamma_{0},\gamma_{1})$ pair   
provided in section 3A [$\Gamma_{3}$: $(6/11,-0.0477)$ and       
$\Gamma_{4}$: $(6/11,0.0113)$].}\label{fig:contours1}
\end{figure}

\subsubsection{Time varying growth index}
Now we concentrate on the $\gamma(z)$ parametrizations,
presented in section 3A. Now the statistical vector becomes: 
${\bf p}=(\Omega_{m0},\gamma_{0},\gamma_{1})$.
Notice that we sample $\gamma_{0} \in [0.1,1.3]$ and
$\gamma_{1} \in [-2.2,2.2]$ in steps of 0.001. 
Since the expansion data put
strong constraints on the value of $\Omega_{m0}$, we find that for 
all $\Gamma_{1-4}$ models, the joint likelihood function peaks 
at $\Omega_{m0}=0.273\pm 0.011$. 

In figures \ref{fig:contours} and \ref{fig:contours1}
we present (after we marginalize over $\Omega_{m0}=0.273$)
the results of our analysis for the $\Gamma_{1}$, 
$\Gamma_{2}$, $\Gamma_{3}$ and $\Gamma_{4}$,
models in the $(\gamma_{0},\gamma_{1})$ plane. 

In the left panels we show the contours using the old growth 
rate data (see Table I, entries 1-7; \cite{Nes08,Gong10,Bas11}) while 
in the right panel 
one can see results for the total sample including that of the 
6dF/{\em WiggleZ}. 
The theoretical $(\gamma_{0},\gamma_{1})$ values in the $\Lambda$CDM 
model indicated by the crosses (see section 3A).

From the left panels of figures 3 and 4, it becomes clear that 
using the old growth rate data-set (entries 1-7)
we are unable to place constraints on the 
$(\gamma_{0},\gamma_{1})$ parameters.
On the other hand, utilizing the overall growth rate sample, we find: 

\begin{table}[ht]
\caption[]{Statistical results for the overall data-set (see Table I): The $1^{st}$ column 
indicates the $\gamma(z)$ parametrizations
 appearing in section 3A.
$2^{nd}$ and $3^{rd}$ columns show the $\gamma_{0}$ and $\gamma_1$ best values.
%The remaining columns present the reduced $\chi^{2}$ and the 
%''reduction factor'' $S$. 
The last column presents the ''reduction factor'' $S$. 
Note that in all case the reduced
$\chi^{2}_{min}$ is $\sim 0.97$.} \tabcolsep 4.5pt
\vspace{1mm}
\begin{tabular}{cccc} \hline \hline
Model& $\gamma_{0}$ & $\gamma_{1}$&$S$ \\ \hline
$\Gamma_{0}$&$0.586^{+0.079}_{-0.074}$& 0 &$2.5$ \\
$\Gamma_{1}$&$0.49^{+0.12}_{-0.11}$& $0.305^{+0.345}_{-0.318}$ &$4.6$ \\
$\Gamma_{2}$&$0.456^{+0.12}_{-0.11}$& $0.587^{+0.502}_{-0.464}$ &$3.9$ \\
$\Gamma_{3}$&$0.461^{+0.12}_{-0.11}$& $0.513^{+0.448}_{-0.414}$ &$3.4$ \\
$\Gamma_{4}$&$0.875^{+0.12}_{-0.11}$& $-0.551^{+0.50}_{-0.46}$&$2.8$ \\ \hline\hline \label{tab:growth1}
\end{tabular}
\end{table}

\noindent
(a) $\Gamma_{1}$ parametrization: 
In this case the likelihood function 
peaks at $\gamma_{0}=0.49^{+0.12}_{-0.11}$ and
$\gamma_{1}=0.305^{+0.345}_{-0.318}$ 
with $\chi_{min}^{2}/dof \simeq 549/567$. 
Interestingly, the addition of five more points (6dF and 
{\em WiggleZ} growth data) 
in the statistical analysis 
provides a significant improvement 
in the derived $(\gamma_{0},\gamma_{1})$ constraints.
In particular, using the overall data-set 
we decrease the $2\sigma$ surface area (see left upper panel of fig.3) 
by a factor of $S \sim 4.6$. 
%Below we will call this quantity ''reduction factor''
%and is indicated by $S$, defined as the ratio of the surface area 
%of the $2\sigma$ contour using the old growth data (entries: 1-7) to that
%of the total growth data-set.
Actually, one would expect such an improvement because the 
6dF and the {\em WiggleZ} surveys measure 
$f(z)$ to within $9-17\%$ in every redshift bin,
in contrast to the old growth rate data 
\cite{Nes08,Gong10,Bas11}
in which the corresponding accuracy lies in the interval $20-40\%$.

\noindent
(b) $\Gamma_{2}$ and $\Gamma_{3}$: Obviously, these 
parametrizations provide similar 
contours and thus they are almost equivalent as far as 
their statistics are concerned. 
%We find that within $1\sigma$
%errors we can put weak constraints on the free parameters. 
In particular, 
the best fit values are: (i) for $\Gamma_{2}$ we have   
$\gamma_{0}=0.456^{+0.11}_{-0.12}$, 
$\gamma_{1}=0.587^{+0.502}_{-0.464}$ and (ii) for 
$\Gamma_{3}$ model we obtain 
$\gamma_{0}=0.461^{+0.12}_{-0.11}$, 
$\gamma_{1}=0.513^{+0.448}_{-0.414}$. In both cases 
the reduced $\chi_{min}^{2}/dof$ is $\simeq 548.4/567$. Notice that 
the ''reduction factor'' here is $S\sim 3.9$ and 3.4 respectively. 

\noindent
(c) $\Gamma_{4}$ parametrization: In this case although 
the $\gamma_{0}$ is strongly degenerate with $\gamma_{1}$, the likelihood 
function peaks at $\gamma_{0}=0.875^{+0.12}_{-0.11}$ and
$\gamma_{1}=-0.551^{+0.50}_{-0.46}$ 
with $\chi_{min}^{2}/dof \simeq 548.4/567$. Also we find that $S\sim 2.8$.

We would like to stress that the predicted 
$(\gamma_{0},\gamma_{1})$
solutions of the $\Gamma_{1-4}$ parametrizations 
remain close to the $1\sigma$ borders (see crosses in figs. 3,4).
In the top panel of figure 5, we present the 
evolution of the growth rate of structure, using the $\Gamma_{0}$ 
parametrization together with the growth data 
scaled to $(\Omega_{m0},\gamma)=(0.273,0.586)$.
In the bottom panel, we present the relative difference between the
$\Gamma_{0}$ parametrization and all the rest $\Gamma_{1-4}$ models, 
ie, $\Delta_{f}(z)=[\Omega^{\gamma(z)}_{m}(z)-\Omega^{0.586}_{m}(z)]/\Omega^{0.586}_{m}(z)]$. The relative growth rate difference of the various fitted $\gamma(z)$ 
models with respect to that of  $\Gamma_{0}$ 
(with $\gamma=0.586$), $\Delta_{f}(z)$ indicates that 
the $\Gamma_{0-4}$ models have a very similar
redshift dependence for $z \ge 0.4$ (with $|\Delta_{f}(z)|\le 0.05$), while 
all the models show large such deviations 
for $z<0.4$, reaching $|\Delta_{f}|\simeq 0.2$ at the lowest redshifts. 

Finally, in Table II, one may see a more compact presentation
of our results for the total sample, including the
''reduction factor'' due to the presence of the 6dF/{\em WiggleZ} data.

\begin{figure}[ht]
\includegraphics[width=0.5\textwidth]{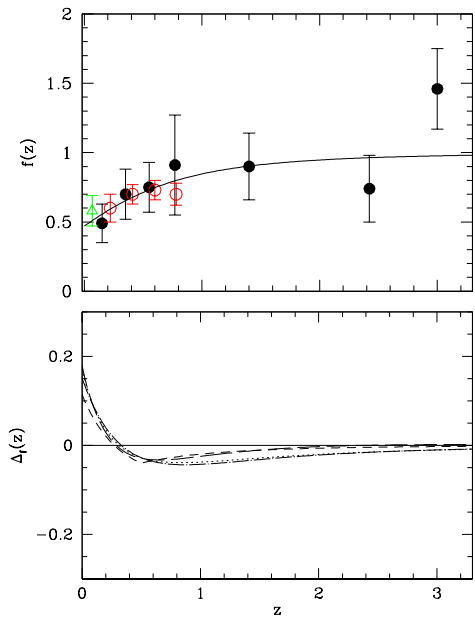}
\caption{{\em Top panel:} Comparison of the observed
and theoretical evolution of the growth
rate of clustering $f(z)=\Omega^{0.586}_{m}(z)$ 
[see solid line: $\Gamma_{0}$ parametrization, $\Omega_{m0}=0.273$]. 
The different growth datasets are
represented by different symbols (see Table I for definitions).
{\em Bottom panel:} The relative difference, $\Delta_{f}(z)$, between
the $\Gamma_{0}$ and the rest of the $\Gamma_{1-4}$ parametrizations.
In particular the different lines correspond to the following pairs:
$\Gamma_{1}-\Gamma_{0}$ (short dashed line),  
$\Gamma_{2}-\Gamma_{0}$ (long dashed line),  
$\Gamma_{3}-\Gamma_{0}$ (dotted line) and $\Gamma_{4}-\Gamma_{0}$ 
(dot-dashed line).   
}\label{fig:growth}
\end{figure}

\section{Conclusions}
In this article we provide a general growth index  
evolution model $\gamma(z)$, based on phenomenology,
which is valid for all possible non-interacting dark energy models, including 
those of modified gravity. 
Armed with our general $\gamma$ evolution model it is straightforward
%to extend the paper of Polarski \& Gannouji \cite{Pol} 
%by applying 
to apply the Polarski \& Gannouji \cite{Pol} 
approach to various $\gamma(z)$ models.
In the context of the concordance $\Lambda$ cosmology, 
we investigate the ability of five growth index 
parametrizations (including a constant one) to 
represent a variety of observational growth 
rate of structure data, based mainly on 2dF, SDSS, VVDS, 6dF and {\em WiggleZ} 
measurements. 
%To this end we perform a $\chi^2$ minimization procedure between
%the observational growth data, after rescaling them to the WMAP7 cosmology, 
%with the model expectations, through which we fit the model
%free parameters. 
To this end we perform a joint likelihood analysis of the
recent expansion data (SNIa, CMB shift parameter and BAOs) 
together with the growth rate of structure data, in order to 
determine the free parameters of the $\gamma(z)$ parametrizations 
and to statistically quantify their ability to represent the observations. 

The comparison shows that all $\gamma$ parametrizations 
fit at an acceptable level the current
growth data, as indicated by the their reduced $\chi^2$ values.
Considering a constant growth index we can place tight constraints, up to
$\sim 15\%$ accuracy, on the $\gamma$ parameter. Indeed, for the total
growth rate data-set (see Table I) 
we find that $\gamma=0.586^{+0.079}_{-0.074}$,
while using only the 6dF and the {\em WiggleZ} growth data 
we obtain $\gamma=0.583^{+0.087}_{-0.082}$,
which is in agreement with 
the theoretically predicted value of $\gamma\simeq 6/11$.
Under the assumption that the growth index varies with time 
we find that the $(\gamma_{0},\gamma_{1})$ parameter solution space 
of all growth index parametrizations, accommodate 
the theoretical $(\gamma_{0},\gamma_{1})$ values at $2\sigma$ level.
We also observe that the inclusion of the new
6dF and {\em WiggleZ} data reduce
significantly the $(\gamma_{0},\gamma_{1})$ parameter solution space. 
Despite the latter improvement we find that the majority of the
$\gamma(z)$ parametrizations still suffer from the 
$\gamma_{0}-\gamma_{1}$ degeneracy, implying that more and accurate
data are essential.

%\vspace {0.4cm}

%{\bf Acknowledgments:}  
%{\it I would like to thank the anonymous referee for useful comments and suggestions.}

\end{document}